\documentclass[12pt,a4paper, reprint, aip, jap, twocolumn, amsmath, amssymb, aps]{revtex4-1} 
\usepackage{graphicx}

\usepackage{floatrow} 
\usepackage[paperwidth=210mm,paperheight=297mm,centering,hmargin=2cm,vmargin=2.1cm]{geometry}
\usepackage{mathptmx}      
\usepackage{amsmath}
\usepackage{setspace}
\usepackage{natbib}

\usepackage{lineno,xcolor}

\begin{document}

\title{Spectral shape deformation in inverse spin Hall voltage in Y$_3$Fe$_5$O$_{12}$$\mid$Pt bilayers at high microwave power levels}

\author{J. Lustikova}
 \email{lustikova@imr.tohoku.ac.jp}
 \affiliation{Institute for Materials Research, Tohoku University, Sendai 980-8577, Japan}

\author{Y. Shiomi}
 \affiliation{Institute for Materials Research, Tohoku University, Sendai 980-8577, Japan}

\author{Y. Handa}
 \affiliation{Institute for Materials Research, Tohoku University, Sendai 980-8577, Japan}

\author{E. Saitoh}
 \affiliation{Institute for Materials Research, Tohoku University, Sendai 980-8577, Japan}
 \affiliation{WPI Advanced Institute for Materials Research, Tohoku University, Sendai 980-8577, Japan}
 \affiliation{CREST, Japan Science and Technology Agency, Tokyo 102-0076, Japan}
 \affiliation{Advanced Science Research Center, Japan Atomic Energy Agency, Tokai 319-1195, Japan}

\date{\today} 
\begin{abstract}
We report on the deformation of microwave absorption spectra and of the inverse spin Hall voltage signals in thin film bilayers of yttrium iron garnet (YIG) and platinum at high microwave power levels in a 9.45-GHz TE$_{011}$ cavity. As the microwave power increases from 0.15 to 200 mW, the resonance field shifts to higher values, and the initially Lorentzian spectra of the microwave absorption intensity as well as the inverse spin Hall voltage signals become asymmetric. The contributions from opening of the magnetization precession cone and heating of YIG cannot well reproduce the data. Control measurements of inverse spin Hall voltages on thin-film YIG$\mid$Pt systems with a range of line widths underscore the role of spin-wave excitations in spectral deformation.
\end{abstract}

\maketitle

\section{Introduction}

Spintronics is a progressive field of electronics, in which in addition to the electronic charge used in conventional electronics, the electron spin is employed for information transmission.\cite{maekawa} In this stream, the generation, manipulation and detection of spin current, the flow of electronic spin angular momentum, are of main interest.

A versatile method for generating spin currents in thin film bilayer systems comprising a ferromagnet and a paramagnetic metal is spin pumping. \cite{silsbee, tserkovnyak-PRL, mizukami, azevedo, costache, saitoh, ando, kajiwara} Upon ferromagnetic resonance (FMR) in the magnetic layer, precession motion of magnetization relaxes not only through damping processes inside the ferromagnet but also by transfer of spin angular momentum to the conduction electrons in the neighbouring paramagnetic layer. 

In a paramagnetic metal with spin orbit coupling, such as platinum, the injected spin current is converted into a transverse charge current by means of the inverse spin Hall effect (ISHE).
\cite{saitoh, valenzuela, kajiwara, ando}
Since the injected spin current is proportional to the power absorbed by the magnetization excitation,\cite{inoue, ando-apl94} spin pumping in combination with ISHE enables direct electric detection of magnetization dynamics in the ferromagnet. 
In the uniform precession mode at low microwave power, the microwave absorption intensity follows a Lorentzian function,\cite{spin-waves,lax} which is reflected by the ISHE voltage signal.\cite{saitoh, ando}

The ferrimagnetic insulator yttrium iron garnet (Y$_3$Fe$_5$O$_{12}$, YIG)\cite{glass} is a commonly used spin injector. The low magnetic loss at microwave frequencies (damping constant $3\times 10^{-5}$)\cite{sparks}, as well as highly insulating properties (band gap 2.85 eV)\cite{sun-wu} make it ideal for the transport and manipulation of pure spin currents. 
YIG also offers a playground for exploring non-linear magnetization phenomena at high microwave power levels, such as spin wave instabilities, foldover and bistable behaviour.\cite{suhl, schloemann}
In recent years, there has been significant interest in observing non-linear spin dynamics in YIG via ISHE measurements and harnessing these effects in spintronics devices.\cite{kurebayashi-nat,kurebayashi-apl, jungfleisch, ando-prl}

The simplest example of a non-linear regime in spin pumping is the effect of decreasing static component of magnetization. \cite{chaos, anderson-suhl, fetisov}
In perpendicular pumping, an external rf field $\textbf{h}_{\text{ac}}$ with frequency $\omega$ is applied perpendicularly to the direction of the static magnetic field $\textbf{H}$, which is pointing along the $z$ direction. 
This causes damped precession motion of the magnetization around the $z$-axis.
The total magnetization vector $\textbf{M}$ then consists of a static component $\textbf{M}_z$ and a dynamic component $\textbf{m}(t)$. For small excitation amplitudes $h_{\text{ac}}$, such that $\mid \textbf{m} \mid  \ll \mid \textbf{M} \mid$, one can approximate  the static component of magnetization by the size of the saturation magnetization, $M_z \approx M_{\rm S}$ (linear regime). With ever increasing microwave power, the decrease in the static component of magnetization $M_z$ can no longer be neglected. This leads to a decrease in the demagnetizing field, which affects the resonance condition.\cite{chaos, spin-waves} 

Such feedback response in the resonance field (or frequency) leads to a "foldover" of the initially Lorentzian microwave absorption spectrum and is responsible for bistability at even higher excitation amplitudes.\cite{prabhakar, fetisov} Foldover and bistability in the FMR of YIG films have been reported previously, and have been attributed either to heat\cite{zhang} or to spin-wave instability processes.\cite{zhang-jap, chen} In Ref. \citenum{iguchil}, nonlinear effects in high power spin pumping have been attributed to spin-wave excitations, but the spectral shape has not been discussed. 

In this work we study the spectral shape deformation of inverse spin Hall voltages in Pt (14 nm) induced by spin pumping from thin YIG films (96 nm). A quantitative analysis of the spectra indicates that they cannot be well explained by the opening of the magnetization precession cone, nor by heating of YIG upon FMR. The disappearance of non-linearity in films with broader line widths suggest that the deformation occurs due to excitations of spin wave modes.

\section{Experiment and results}

YIG films investigated in this study were deposited by on-axis magnetron rf sputtering on gadolinium gallium garnet (111) (Gd$_3$Ga$_5$O$_{12}$, GGG) substrates with a thickness of 500 $\mu$m. The base pressure was $2.3\times 10^{-5}$ Pa, and the pressure of the pure argon atmosphere was $1.3$ Pa. During deposition, the substrate remained at ambient temperature and the deposition rate was $2.7$ nm/min. Crystalization was realized by post-annealing in air at $850$ $^{\circ}$C for 24 hours. Thickness of the films was 96 nanometers. The effective saturation magnetization and the damping constant of the films were $103\pm 5$ kA/m and $(7.0\pm 1.0)\times10^{-4}$, respectively; with a peak-to-peak line width of $(0.40\pm0.03)$ mT at a frequency of $9.45$ GHz. The structural properties as well as the spin injection efficiency in the linear low-power regime of such films have been described elsewhere.\cite{lustikova}

For spin current detection, the YIG samples were coated by a 14-nm-thick platinum film by rf sputtering, causing the line width to enhance to $(0.52\pm0.02)$ mT. Platinum was chosen for its high conversion efficiency from spin current to charge current.\cite{ando} 
The width and length of the samples were $w=1$ mm and $l=3$ mm, respectively.

\begin{figure}[t]
\centerline{\includegraphics[width=8.5cm, angle=0, bb=0 0 450 480]{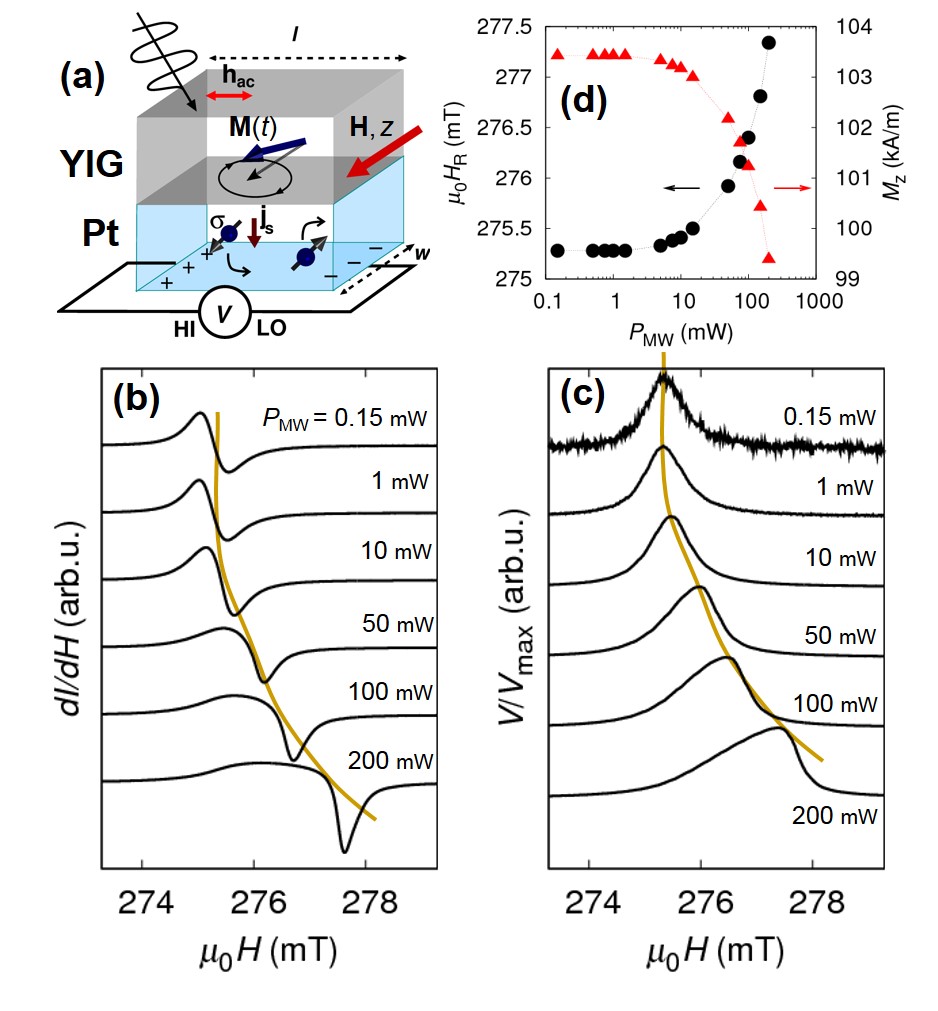}}
\caption{\label{fig1}
(a) Schematic illustration of the experimental setup. $\mathbf{H}$, $\mathbf{h}_{\text{ac}}$, $\mathbf{M}(t)$, $\mathbf{j}_{\rm s}$ and $\boldsymbol{\sigma}$ denote the static magnetic field, the rf field, the magnetization vector, the injection direction of spin current generated by spin pumping and the spin-polarization vector of the spin current, respectively. The bent arrows in the Pt layer illustrate the motion of the electrons under the influence of the spin-orbit coupling which leads to the appearance of a transverse electromotive force (ISHE).
(b) Field $H$ sweep of the microwave absorption intensity differentiated by $H$, 
(c) normalized inverse spin Hall voltage at selected values of microwave power $P_{\text{MW}}$. Here, $V_{\text{max}}$ is the peak value of the voltage signal. 
The yellow curves are eye guides visualising the position of the resonance at each $P_{\text{MW}}$.
(d) $P_{\text{MW}}$ dependence of the observed ferromagnetic resonance field $H_{\text R}$ (black circles) and of the static component of the magnetization $M_z$ (red triangles) determined from the resonance condition, Eq. \eqref{FMRcond}.
}
\end{figure}

Figure \ref{fig1} summarizes the setup and the results of the spin pumping experiment on samples prepared by sputtering. Figure \ref{fig1}(a) is an illustration of the experimental setup. The measurement was performed at room temperature in a 9.45-GHz TE$_{011}$ cylindrical microwave cavity with microwave power $P_{\text{MW}}$ in the range from 0.15 mW (corresponding to an rf field $\mu_0 h_{\text{ac}}=4.4$ $\mu$T) to 200 mW ($\mu_0 h_{\text{ac}}=0.16$ mT). The sample was placed in the centre of the cavity where the electric field component of the microwave is minimized while the magnetic field component is maximized and lies in the plane of the sample surface. A static magnetic field was applied in the plane of the sample surface perpendicular to the direction of the rf field and to the direction in which the ISHE voltage was measured. The microwave absorption intensity was measured using a field lock-in technique.\cite{ando}

The obtained microwave absorption derivative spectra $dI/dH$ and normalized ISHE voltage signals $V/V_{\text{max}}$ at various values of $P_{\text{MW}}$ are shown in Fig. \ref{fig1} (b) and (c), respectively. Here, $V_{\text{max}}$ denotes the maximal value of the voltage peak. All curves were obtained by sweeping the static field at a fixed rate of 60 mT/min.

At the lowest power ($P_{\text{MW}}=0.15$ mW), the microwave absorption derivative spectrum $dI/dH$ in Fig. \ref{fig1}(b) has the shape of the first derivative of a Lorentzian function, as predicted by linear magnetization dynamics. The corresponding ISHE voltage signal in Fig. \ref{fig1}(c) also follows a Lorentzian profile, with the same resonance field $H_{\rm R}$ and full width at half maximum as the microwave absorption spectrum. 
The ISHE origin of the voltage signal has been confirmed in Ref. \citenum{lustikova}. 
This Lorentzian behaviour is preserved in the low power regime ($P_{\text{MW}}=0.15$, 1, 10 mW).

As the microwave power is increased beyond 10 mW ($P_{\text{MW}}=50$, 100, 200 mW) a shift of the resonance field $H_{\rm R}$ to higher values occurs. Along with this $H_{\rm R}$ shift,  the microwave absorption spectra as well as the voltage signals gradually develop a deformed shape. This feature is most pronounced at the highest power, $P_\text{MW}=200$ mW. Here, the left shoulder of the microwave absorption derivative is significantly broader than the right one, which exhibits a sharp dip. This derivative spectrum shows good correspondence with the voltage signal, which has the shape of an inclined Lorentzian peak with a broad left shoulder and a narrow right shoulder. Subsidiary resonance peaks were not observed in the spin pumping measurement.

The shift of the measured resonance field $\mu_0 H_{\text{R}}$ to higher values with increasing $P_{\text{MW}}$ is shown in Fig. \ref{fig1}(d). As the power is increased from 0.15 mW to 200 mW, $\mu_0 H_{\text{R}}$ is first constant at a value of 275.3 mT up to $P_{\text{MW}}=8$ mW, and after that, gradually increases to 277.3 mT at the highest microwave power.

This increase in resonance field $H_{\rm R}$ points to a decrease in the static component of the magnetization $M_z$ which can be estimated from the resonance condition.\cite{zhang, zhang-jap}
For tangentially magnetized films (with static magnetic field in the $z$ -direction), an rf demagnetizing field is created by the out-of plane dynamic component of magnetization, leading to a resonance condition\cite{spin-waves}

\begin{equation}
\left( \frac{\omega}{\gamma}\right)^2=\mu_0 H_{\text{R}}\left( \mu_0 H_{\rm R}+\mu_0 M_z\right),
\label{FMRcond}
\end{equation}

based on which the observed $H_{\rm R}$ shift corresponds to a decrease in $M_z$ from 103.4 kA/m to 99.4 kA/m. The $P_{\text{MW}}$ dependence of $M_z$ is plotted in Fig. \ref{fig1}(d). Reflecting the behaviour of $H_{\rm R}$, the static component of magnetization $M_z$ is first constant up to $P_\text{MW}=8$ mW, and after that gradually decreases with increasing microwave power.


\section{Analysis and discussion}

The observation of asymmetric spectral profiles in the microwave absorption intensity and of those in the ISHE voltage is likely linked to a decrease in the static component of the effective saturation magnetization, which can be caused by (i) opening of the precession cone, (ii) heating of the ferromagnet, or (iii) spin wave instability processes. \cite{zhang-jap} We first examine the uniform precession mode, case (i) and (ii).

\begin{figure}[b]
\centerline{\includegraphics[width=8.5cm, angle=0, bb=25 15 580 400]{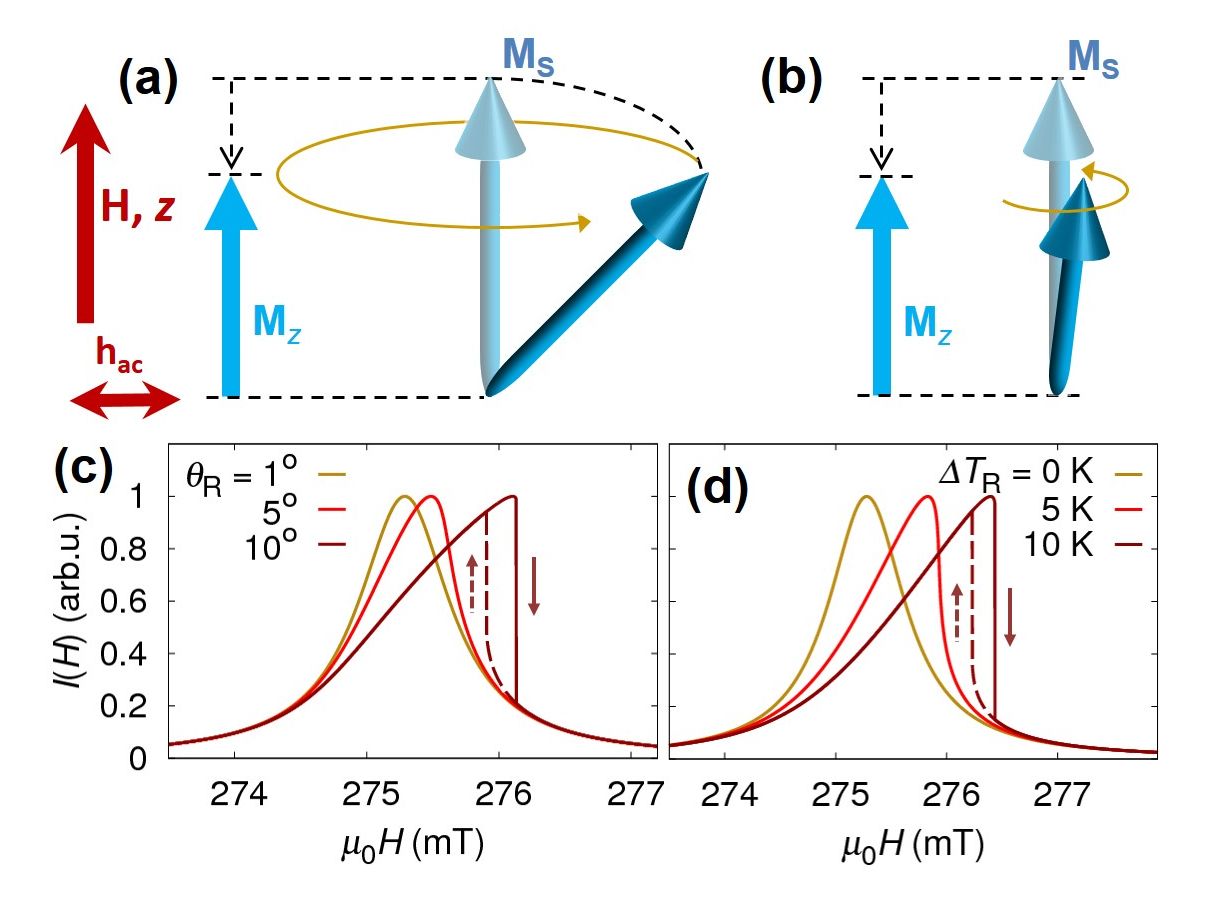}}
\caption{\label{fig2}
 Two possible mechanisms of spectral shape deformation at high microwave power levels. 
(a) Opening of the precession cone causes the $z$-component of the magnetization vector (blue arrows) to decrease from the saturation value $M_{\text{S}}$ to a smaller value $M_z$. (b) The magnetization vector of the ferromagnet ``shrinks" due to thermal fluctuations caused by heating upon FMR, leading to a smaller $z$-component $M_z$.
 Results of the numerical calculation of spectral shape of the microwave absorption intensity using Eq. \eqref{newL}, (c) for the opening of the precession cone at selected values of precession angle $\theta_{\rm R}$ at FMR, and (d) for heating at selected values of temperature increase  $\Delta T_{\rm R}$ at FMR.}
\end{figure}

These two mechanisms are illustrated in Figs. \ref{fig2}(a) and (b). In Fig. \ref{fig2}(a), increasing the rf field leads to the opening of the precession cone. While the total magnetization vector $\textbf{M}$ remains unchanged, the static component $M_z$, which is the projection of $\textbf{M}$ into the direction of the external field, decreases. In Fig. \ref{fig2}(b), the heating of the ferromagnet due to microwave absorption is considered. The increasing thermal fluctuations of the individual spins result into a decrease in the total magnetization $\textbf{M}$, leading to a smaller static component $M_z$.

\begin{figure*}
\centerline{\includegraphics[width=16cm, angle=0, bb= 0 0 1000 250]{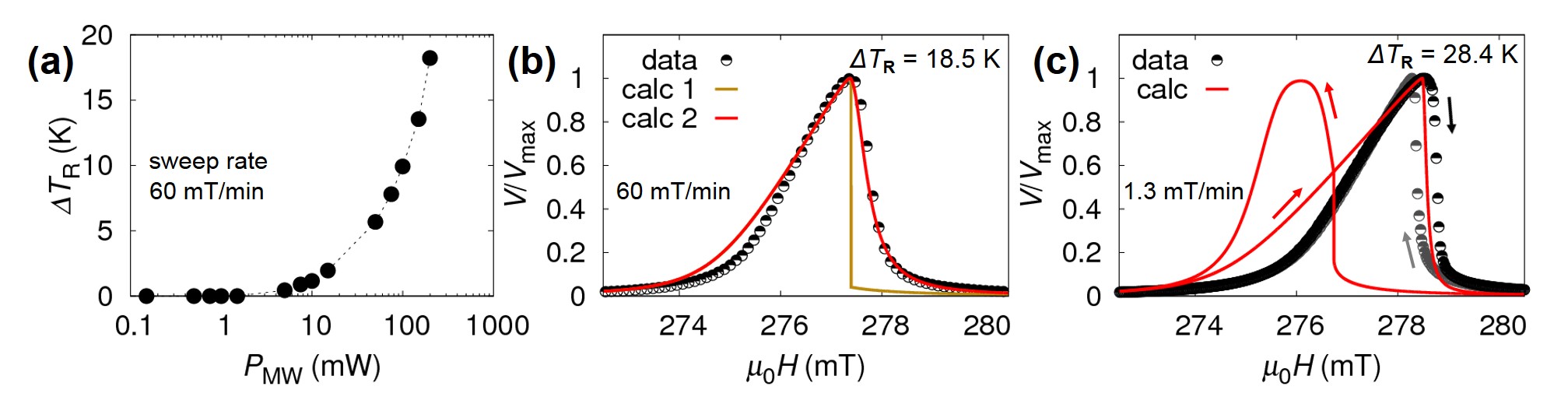}}
\caption{\label{fig3}
(a) Temperature increase $\Delta T_{\rm R}$ calculated from the ferromagnetic resonance field $H_{\rm R}$ at each $P_{\text{MW}}$ and Eq. \eqref{temp}.
(b) Comparison of the ISHE voltage signal observed at $P_{\text{MW}}=200$ mW and a sweep rate of 60 mT/min (``data") and the numerical calculation for $\Delta T$ proportional to $I(H)$ at every point of the sweep corresponding to immediate cooling (``calc 1") and for a relaxation time of $\tau= 15$ s (``calc 2").
(c) Comparison of the ISHE voltage signal observed at $P_{\text{MW}}=200$ mW and a sweep rate of 1.3 mT/min in upward (black dots) and downward sweep (grey dots) with calculated curves for $\tau=40$ s (``calc"), for upward and downward sweep as indicated by red arrows.
}
\end{figure*}

In both cases, the increase in microwave absorption intensity with increasing power leads to a decrease in the static component of magnetization $M_z$, which manifests itself as an increase in the resonance field $H_{\text{R}}$ via the resonance condition in Eq. \eqref{FMRcond}. One then has to consider the dependence of $H_{\text{R}}$ on the microwave absorption intensity $I$, so that the originally Lorentzian spectral shape of $I(H)$ assumes the following form:\cite{prabhakar}

\begin{equation}
I(H)=\frac{\Gamma^2}{\left(H-H_{\rm R}(I)\right)^2+\Gamma^2}, \label{newL}
\end{equation}
where $\Gamma$ is a damping factor and the expression is normalized so that $I(H_{\text{R}})=1$. 
The dependence of the resonance field on the microwave absorption intensity, $H_{\rm R}=H_{\rm R}(I)$, causes a deformation of the microwave absorption spectrum $I(H)$ into an inclined Lorentzian peak.

At each point of the $H$-sweep, $H_{\text{R}}$ is determined from Eq. \eqref{FMRcond} as 
$\mu_0 H_{\rm R}=-\frac{1}{2} \mu_0 M_z +\frac{1}{2} \sqrt{(\mu_0 M_z)^2 +4\left( \omega/\gamma\right)^2}$. 
The spectral shape of the microwave absorption intensity can be then obtained by solving Eq. \eqref{newL} for a given dependence of $M_z$ on the microwave absorption intensity $I(H)$. In the analysis below we fix the damping parameter $\Gamma=0.42$ mT, as determined from Lorentzian fit of the data at $P_{\text{MW}}=0.15$ mW, and use $(\omega/\gamma)=0.334$ T.

We first analyze the case of the opening precession cone [case (i)]. The precession angle $\theta$ is defined as the angle between the magnetization vector $\textbf{M}$ and the direction of the static magnetic field ($z$-direction), so that $M_z=M_{\rm S} \cos \theta$. We assume that the precession angle $\theta$ increases linearly with the microwave absorption intensity, $\theta=k_1 I$. The normalization requirement $I(H_{\rm R})=1$ leads to a coefficient $k_1=\theta_{\rm R}$, where $\theta_{\rm R}$ is the precession angle at FMR. We obtain the spectral shape of the microwave absorption intensity, $I(H)$, by solving Eq. \eqref{newL} with $H_{\rm R}(I)$ determined from Eq. \eqref{FMRcond}, where $M_z(I)=M_{\rm S} \cos(\theta_{\rm R} I)$. The saturation magnetization is here fixed at $M_{\rm S}=103$ kA/m as determined from the FMR condition at $P_{\text{MW}}=0.15$ mW.

The numerical solutions of Eq. \eqref{newL} for selected values of $\theta_{\text{R}}$ are plotted in Fig. \ref{fig2}(c). With increasing $\theta_{\rm R}$, the spectra develop an asymmetric shape with $H_{\rm R}$ shifting towards higher values. At $\theta_{\rm R}=10^\circ$, Eq. \eqref{newL} has more than one solution, which leads to different values of microwave absorption intensity when sweeping the magnetic field in opposite directions. When sweeping the static magnetic field in the upward direction, $I(H)$ increases up to a maximum at $H_1$, after which is suddenly drops. For a sweep in the downward direction, $I(H)$ increases only slightly up to $H_2$ (smaller than $H_1$), where it abruptly jumps to its value at the upward sweep, and then decreases along the same path as the in upward sweep.

This hysteretic behaviour of the numerical solution sets in at $\theta_{\rm R}=8^\circ$, at which the resonance field predicted from $M_z$ by Eq. \eqref{FMRcond} is $H_{\rm R}= 275.8$ mT. In our experiments, this value of $H_{\rm R}$ was observed for $P_{\text{MW}}$ between 10 mW and 50 mW where the experimental spectra are nowhere near hysteretic. Therefore, the observed spectra cannot be explained by the opening of the precession cone.

Next, we discuss heating of the YIG upon FMR [case (ii)]. Here, we assume the precession angle to be negligibly small so that $M_z\approx M_{\rm S}$.
According to magnetization measurements, the temperature dependence of the magnetization of YIG in the vicinity of $300$ K is approximately linear,\cite{anderson} and can be expressed as

\begin{equation}
M_z(T)=M_{\text{S}}(1-\kappa \Delta T), \label{temp}
\end{equation}
where $M_{\text{S}}$ is the effective saturation magnetization at $300$ K, and $\Delta T$ is the increase in the temperature of YIG, with base taken at 300 K.
The linear coefficient was estimated as $\kappa=2.14 \times 10^{-3}$ K$^{-1}$ based on Ref. \citenum{anderson}  by a linear fit of the magnetization curve [Fig. 1 in Ref. \citenum{anderson}] in the vicinity of 300 K. The coefficient in our samples, obtained by measuring the temperature dependence of magnetization, was $2.23\times 10^{-3}$ K$^{-1}$.

We assume that $\Delta T$ is proportional to the absorbed microwave power $I(H)$ at every point of the field sweep, $\Delta T(H)= k_2 I(H).$ Again, due to the normalization $I(H_{\text{R}})=1$, we have $k_2=\Delta T_{\rm R}$, where $\Delta T_{\rm R}$ denotes the temperature increase at FMR. 
 To obtain the spectral shape of the microwave absorption intensity $I(H)$, we solve Eq. \eqref{newL} with $H_{\rm R}$ determined from Eq. \eqref{FMRcond}, where $M_z(T)$ is given by Eq. \eqref{temp}. 
In the calculation, we take $M_{\text S}=103$ kA/m.

The calculated curve of the FMR spectrum for selected values of $\Delta T_{\rm R}$ is plotted in Fig. \ref{fig2}(d). As the temperature increases, the spectrum takes on an asymmetric shape with the peak shifting towards higher values of magnetic field. The left shoulder of the FMR peak broadens with increasing temperature. The apparent line width of the FMR peak increases with increasing $\Delta T_{\rm R}$. For $\Delta T_{\text{R}}=10$ K we observe hysteretic behaviour between the upward and downward field sweep.

Based on Eq. \eqref{temp} we first estimate the temperature increase $\Delta T_{\text{R}}$ at ferromagnetic resonance that would be necessary to explain the magnitude of the decrease in $M_z$ with increasing microwave power shown in Fig. \ref{fig1}(d). The result is shown in Fig. \ref{fig3}(a). Up to $P_{\rm{MW}}=8$ mW, $\Delta T_{\rm R}$ is zero, and after that gradually increases with increasing microwave power. The overall behaviour reflects the $P_{\text{MW}}$ dependence of $H_{\rm R}$ and $M_z$. For $P_{\rm{MW}}=200$ mW, a temperature increase of $18.5$ K is estimated.

Figures \ref{fig3}(b), (c) compare the results of the numerical calculation for case (ii) with the voltage signal observed at $P_{\text{MW}}=200$ mW. Here, the assumption is that the ISHE voltage observed in the spin pumping is directly proportional to the absorbed microwave power $I(H)$.

In Fig. \ref{fig3}(b) the ISHE voltage obtained at $P_{\text{MW}}=200$ mW and a sweep rate of 60 mT/min is compared with the result of the calculation for $\Delta T_{\text{R}}=18.5$ K (``calc 1"). While the voltage signal is smooth, the calculated curve drops sharply after resonance and is a bad fit to the experimental data. To obtain a good fit at the right flank, it is necessary to assume that the temperature after resonance decreases exponentially back to 300 K with a relaxation constant $\tau$, that is, $\Delta T(H)= \Delta T_{\rm R} \exp[-(H-H_{\rm R})/(v\cdot\tau)]$. Here, $v$ is the field-sweep rate. The best agreement between the signal and the calculated curve  was obtained for $\tau= 15$ s (``calc 2") which signifies a rather slow cooling process. However, there is a slight disagreement between data and calculation at the left flank.

In Fig. \ref{fig3}(c) we show the ISHE signal obtained at $P_{\text{MW}}=200$ mW and a sweep rate of 1.3 mT/min. The shape of the signal resembles that at higher sweep rate, but $H_{\text{R}}$ is shifted to higher values (278.5 mT) leading to an estimate $\Delta T_{\text R}=28.4$ K. Hysteretic behaviour was observed, namely, the resonance field in the downward sweep (278.3 mT) is smaller than in the upward sweep (278.5 mT). Calculation for $\Delta T_{\text R}=28.4$ K is plotted along with the data.
The disagreement between the left flank of the signal and the calculation in the upward sweep is even larger than in Fig. \ref{fig3}(b).
Further, to simulate the slow decrease of the voltage at the right flank in the upward sweep, a relaxation time of more than $\tau=40$ s is required.  Moreover, the calculated curve in the downward sweep reaches resonance at 276 mT which is in strong disagreement with the observed curve.

There are several problems with the heating model. The cooling times estimated from the experimental curves at 60 mT/min and 1.3 mT/min, $\tau=15$ s and more than 40 s, respectively, are unreasonably high considering that the small volume of the ferromagnetic film, attached to a thick GGG substrate acting as a heat bath, should cool almost instantaneously. In addition, the modelled curves do not match the left flank of the signal at any sweep rate. 
Finally, the temperature increase $\Delta T_{\text{R}}$ required to explain the $H_{\text{R}}$ shift leads to a hysteresis that is much larger than the observed one; namely a $H_{\text{R}}$ difference between upward and downward sweep that is almost 8 times higher than the observed one for the $1.3$ mT/min sweep [Fig. \ref{fig3}(c)]. From these observations we conclude that a simple heating model cannot explain the data. 

The discussion above suggests that the origin of the non-linear phenomena observed is neither opening of the precession cone, nor heating upon FMR. A possible mechanism of the decrease in the resonance field $H_{\text{R}}$ with increasing microwave field might be the outflow of $M_z$ component into spin waves with zero or non-zero wave vector. Consequently, the deformation of the FMR spectra as well as the ISHE signals might be related to the excitation of spin waves.  

\section{Control measurements on YIG films with different line widths}

To further investigate the behaviour at high power microwave levels, we have performed spin pumping from sputtered YIG films which, in addition to annealing in air, have also been annealed in vacuum, as well as from films prepared by pulsed laser deposition (PLD).

Films prepared by sputtering under the conditions described in section II., were additionally annealed in vacuum at 500 $^\circ$C for 3 hours.
The line width of the samples at $9.45$ GHz increased from $(0.95\pm0.21)$ mT to $(2.14\pm0.13)$ mT by this process.
A possible reason for the line width enhancement are oxygen vacancies near the sample surface introduced during the vacuum annealing process.

\begin{figure}[t]
\centerline{\includegraphics[width=8.5cm, angle=0, bb=20 0 670 260]{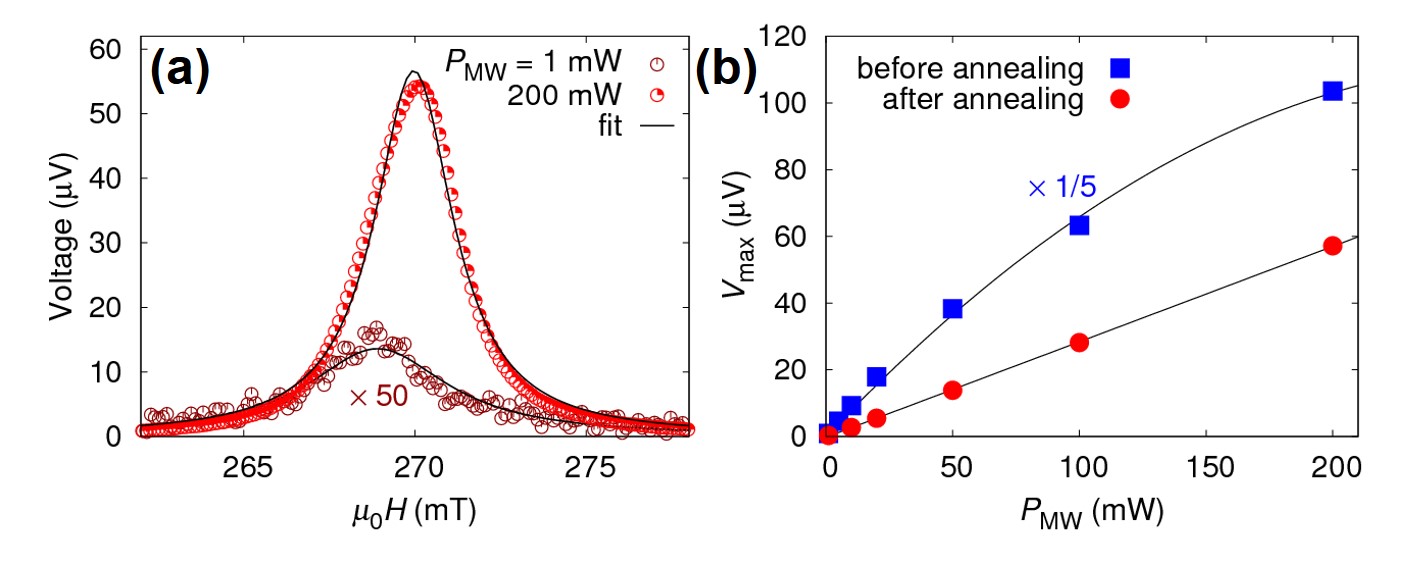}}
\caption{\label{fig4}
(a) Inverse spin Hall voltage signal observed in a YIG(96 nm)$\mid$Pt(14 nm) sample where the YIG was additionally annealed in vacuum prior to Pt coating. The signal at $P_{\text{MW}}=1$ mW is plotted 50 times larger for increased visibility. The fit is a Lorentzian function.
(b) Microwave power dependence of the peak value of the ISHE voltage observed in a YIG$\mid$Pt sample which has been coated by Pt directly after annealing of YIG in air ("before annealing") and which has in addition been  annealed in vacuum before Pt deposition ("after annealing"). The signal from the former sample is plotted reduced by a factor 5 for improved visibility. The black lines are a guides for the eyes.
}
\end{figure}

Figure \ref{fig4} presents the effects of the vacuum annealing on spin pumping. In Fig. \ref{fig4}(a) we show the inverse spin Hall voltage signal from a sample where the YIG has been annealed in air, later in vacuum, and finally coated by 14-nm Pt layer by sputtering at room temperature. The ISHE signal at both low ($P_{\text{MW}}=1$ mW) and high ($P_{\text{MW}}=200 $ mW) microwave power has a Lorentzian shape. The spectral deformation is neither present in the microwave absorption spectra. Fig. \ref{fig4}(b) shows the $P_{\text{MW}}$ dependence of the ISHE voltage at FMR in a sample where the YIG has only been annealed in air ("before annealing") and a sample where the YIG has been annealed in air and additionally in vacuum ("after annealing"). In the former case, the power dependence of ISHE voltage deviates from the predicted linear dependence, namely, the signal amplitude at higher power levels is smaller than that given by the expected linear dependence. However, in the latter case the power dependence follows a straight line as expected in the conventional spin pumping model.\cite{ando} Annealing of the YIG in vacuum caused the magnitude of the ISHE signal at $P_{\text{MW}}=200$ mW become smaller by a factor of 9 compared to untreated samples.

Finally, we look at spin pumping in a YIG$\mid$Pt system where the YIG has been prepared by PLD.  The films were deposited on GGG substrates from a stoichiometric target using a KrF excimer laser with a repetition rate of 10 Hz. During the deposition, the GGG substrate was kept at 750 $^\circ$C and a pure oxygen atmosphere with a pressure of 27 Pa was maintained. The deposition rate was 0.053 nm/min and the final thickness 9 nm. After the growth, the films were annealed at $800$ $^\circ$C in 63 kPa oxygen gas for 1 hour.

\begin{figure}[b]
\centerline{\includegraphics[width=8.5cm, angle=0, bb= 20 0 600 480]{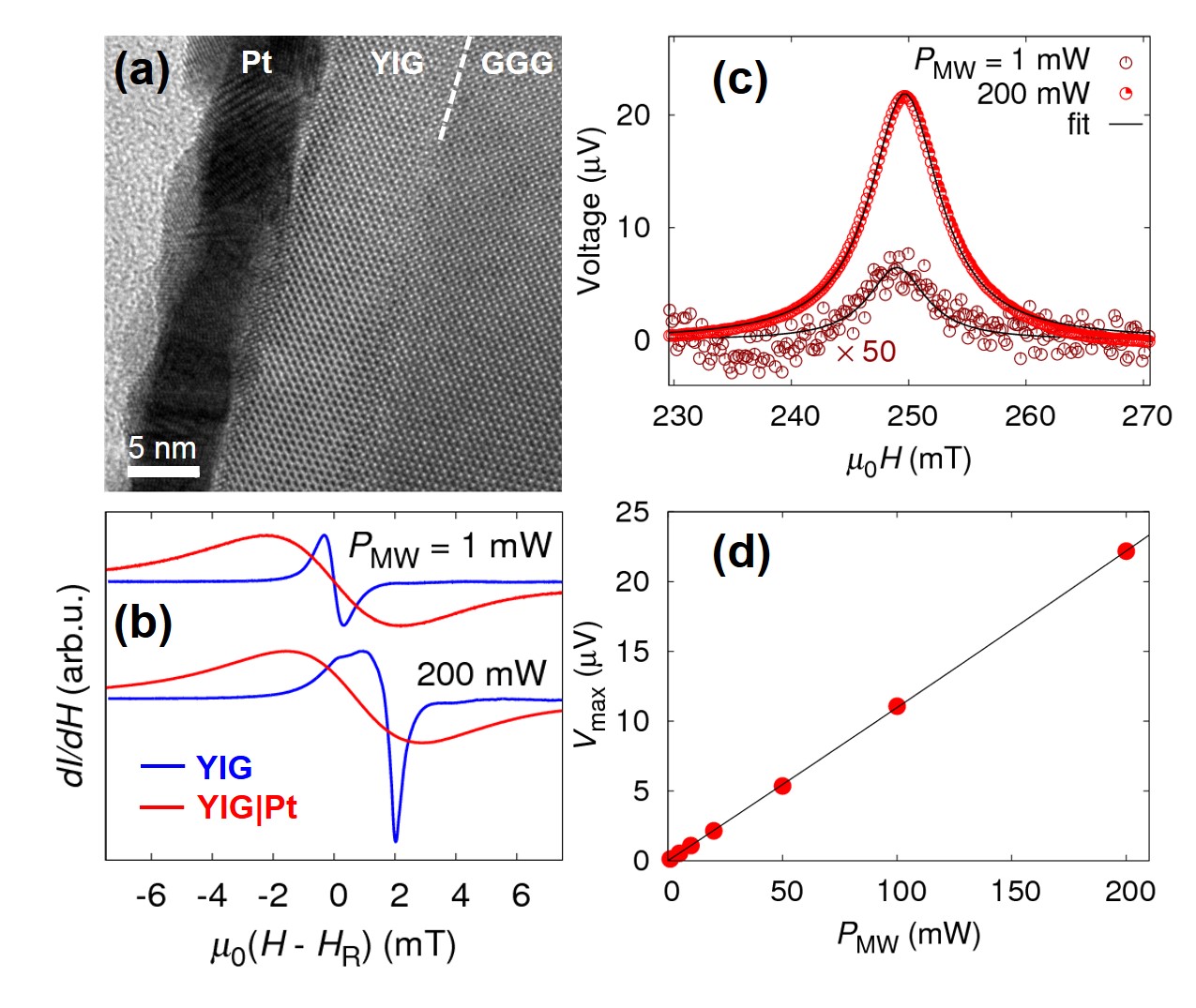}}
\caption{\label{fig5}
Spin pumping from YIG prepared by PLD.
(a) Cross-section TEM image of a GGG$\mid$YIG(9 nm)$\mid$Pt(8 nm) sample.
(b) FMR spectra of a bare YIG(9 nm) sample (blue curves) and a YIG(9 nm)$\mid$Pt(14 nm) sample (red curves) at $P_{\text{MW}}=1$ mW and $200$ mW. Here, $H_{\text{R}}$ denotes the resonance field at $P_{\text{MW}}=1$ mW.
(c) Inverse spin Hall voltage signal in a YIG(9 nm)$\mid$Pt(14 nm) sample at $P_{\text{MW}}=1$ mW and $200$ mW. The signal at $P_{\text{MW}}=1$ mW has been plotted multiplied by factor 50 to improve visibility. The fit is a Lorentzian function.
(d) Microwave power dependence of the peak value of the ISHE voltage (red points). The black line is a guide for the eyes.
}
\end{figure}

 The results are summarized in Fig. \ref{fig5}. The TEM cross-section image of a GGG$\mid$YIG(9 nm)$\mid$Pt(8 nm) sample is shown in Fig. \ref{fig5}(a). The YIG grows epitaxially on the GGG substrate and the garnet structure is perfectly maintained throughout the thin film. For spin pumping experiments, the YIG has been coated by 14-nm Pt layer by sputtering. A comparison of the microwave absorption derivative spectra prior to and after Pt sputtering is shown in Fig. \ref{fig5}(b). Prior to Pt coating (``YIG"), the spectrum is a Lorentzian function derivative for low microwave power ($P_{\text{MW}}=1$ mW). At $P_{\text{MW}}=200$ mW a deformation of the spectrum and an $H_{\text{R}}$ shift qualitatively similar to those presented in Fig. \ref{fig1}(b) were observed. After Pt coating (``YIG$\mid$Pt"), the microwave absorption spectrum broadens, has a Lorentzian derivative shape and does not change by increasing the power from $P_{\text{MW}}=1$ mW to $200$ mW. The spectral width enhancement measured at $P_{\text{MW}}=1$ mW was from $0.64$ mT in bare YIG to $4.40$ mT in YIG$\mid$Pt at a frequency of $9.45$ GHz. The shift of $H_{\text{R}}$ by increasing the power from $P_{\text{MW}}=1$ mW to $200$ mW corresponds to a decrease in effective $M_z$ from $142$ kA/m to $136$ kA/m in YIG, and from $162$ kA/m to $160$ kA/m in YIG$\mid$Pt.

The results of the ISHE measurements on the (PLD-made-YIG)$\mid$Pt system are shown in Figs. \ref{fig5}(c),(d). The spectral shape was that of a Lorentzian for all microwave powers from $P_{\text{MW}}=1$ mW to $200$ mW [Fig. \ref{fig5}(c)]. The power dependence of the ISHE signal amplitude is linear [Fig. \ref{fig5}(d)].

The overall trends observed in the measurements on YIG$\mid$Pt systems where the YIG has been prepared by (A) sputtering at room temperature and subsequent post-annealing in air, (B) same as (A) and additional annealing in vacuum, (C) pulsed laser deposition, are the following. (i) A "foldover" of the microwave absorption spectra as well as of the inverse spin Hall voltage signals with increasing microwave power was observed in films with small line widths (below $1$ mT at $9.45$ GHz). The spectra remained Lorentzian in films with broader line widths (more than $2$ mT at $9.45$ GHz). (ii) At increased microwave power levels, a shift of the resonance field corresponding to a decreased effective static magnetization component develops. (iii) A strongly non-linear power dependence of the ISHE voltage amplitude was observed along with the spectral shape deformation. The power dependence in films with Lorentzian spectra was linear even at high microwave power levels.

The fact that the deformation of spectral shapes, as well as the non-linear power dependence of the ISHE voltages are only present in films with lower damping may be explained in terms of spin-wave excitations, which can cause a reduction of $M_z$. Spin pumping is realized for spin waves with zero as well as non-zero wave vectors, and these spin-wave excitations can be detected via ISHE.\cite{jungfleisch,sandweg}
A similar non-linear power dependence of the ISHE voltage attributed to spin-wave excitations has been observed in 200-nm thick YIG  with a damping constant $2\times10^{-4}$ which was prepared by liquid phase epitaxy,\cite{castel} and demonstrated even in spin pumping from Bi:YIG films prepared by metal-organic decomposition method.\cite{iguchil}

\section{Conclusion}

In summary, we have investigated the microwave absorption spectra of tangentially magnetized 96-nm thick sputtered YIG films and the inverse spin Hall voltage signals induced in adjacent Pt layers by spin pumping in a TE$_{011}$ microwave cavity. We have found that with increasing microwave power, the absorption spectra as well as the voltage signals develop an asymmetric shape with a broadened left shoulder and that the resonance field shifts to higher values, which points to a decrease in the static component of the magnetization.  Analysis of the spectral shapes and comparison with measurements on other YIG$\mid$Pt systems suggests that this deformation may be caused by spin-wave excitation processes. Inverse spin Hall effect enables direct electrical detection of this process.

\begin{acknowledgments}
This work was supported by 
CREST ``Creation of Nanosystems with Novel Functions
through Process Integration", 
Strategic International Cooperative Program ASPIMATT from JST, Japan, and 
Grants-in-Aid for Challenging Exploratory Research (No. 26610091) and 
Scientific Research (A) (No. 24244051) from MEXT, Japan.
\end{acknowledgments}


\begin{thebibliography}{00}
\addcontentsline{toc}{chapter}{References}

\bibitem{maekawa} S. Maekawa, \textit{Concepts in Spin Electronics} (Oxford Univ. Press, 2006).

\bibitem{silsbee} R. H. Silsbee, A. Janossy, and P. Monod, Phys. Rev. B \textbf{19}, 4382 (1979).
\bibitem{tserkovnyak-PRL} Y. Tserkovnyak, A. Brataas, and G. E. W. Bauer, Phys. Rev. Lett. \textbf{88}, 117601 (2002).
\bibitem{mizukami} S. Mizukami, Y. Ando, and T. Miyazaki, Phys. Rev. B \textbf{66}, 104413 (2002). 
\bibitem{azevedo} A. Azevedo, L. H. Vilela Leao, R. L. Rodriguez-Suarez, A. B. Oliveira, and S. M. Rezende, J. Appl. Phys. \textbf{97}, 10C715 (2005).
\bibitem{saitoh} E. Saitoh, M. Ueda, H. Miyajima, and G. Tatara, Appl. Phys. Lett. \textbf{88}, 182509 (2006).1
\bibitem{costache} M. V. Costache, M. Sladkov, S. M. Watts, C. H. van der Wal, and B. J. van Wees, Phys. Rev. Lett. \textbf{97}, 216603 (2006).
\bibitem{kajiwara} Y. Kajiwara, K. Harii, S. Takahashi, J. Ohe, K. Uchida, M. Mizuguchi, H. Umezawa, H. Kawai, K. Ando, K. Takanashi, S. Maekawa, and E. Saitoh, Nature \textbf{464}, 262 (2010).
\bibitem{ando} K. Ando, S. Takahashi, J. Ieda, Y. Kajiwara, H. Nakayama, T. Yoshino, K. Harii, Y. Fujikawa, M. Matsuo, S. Maekawa, and E. Saitoh, J. Appl. Phys. \textbf{109}, 103913 (2011).

\bibitem{valenzuela} S. O. Valenzuela and M. Tinkham, Nature \textbf{442}, 176 (2006). 

\bibitem{inoue} H. Y. Inoue, K. Harii, K. Ando. K. Sasage, and E. Saitoh, J. Appl. Phys. \textbf{102}, 083915 (2007).
\bibitem{ando-apl94} K. Ando, J. Ieda, K. Sasage, S. Takahashi, S. Maekawa, and E. Saitoh, Appl. Phys. Lett. \textbf{94}, 262505 (2009).


\bibitem{spin-waves} D. D. Stancil, and A. Prabhakar, $Spin$ $Waves$ (Springer, 2009).
\bibitem{lax} B. Lax, and K. J. Button, \textit{Microwave Ferrites and Ferrimagnetics} (McGraw Hill, 1962).


\bibitem{glass} H. L. Glass, Proc. IEEE \textbf{76}, 151 (1988).

\bibitem{sparks} M. Sparks, $Ferromagnetic$ $relaxation$ $theory$ (McGraw Hill, New York, 1964).
\bibitem{sun-wu} Y. Sun and M. Wu, in Solid State Physics, edited by M. Wu and A. Hoffmann (Elsevier, 2014), Vol. 64, Chap. 6.

\bibitem{suhl} H. Suhl, J. Phys. Chem. Solids \textbf{1}, 209 (1957).
\bibitem{schloemann} E. Schl\"omann, J. J. Green, and U. Milano, J. Appl. Phys. \textbf{31}, S386 (1960).

\bibitem{kurebayashi-nat} H. Kurebayashi, O. Dzyapko, V. E. Demidov, D. Fang, A. J. Fergusson, and S. O. Demokritov, Nature Mat. \textbf{10}, 660 (2011).
\bibitem{kurebayashi-apl} H. Kurebayashi, O. Dzyapko V. E. Demidov, D. Fang, A. J. Ferguson, and S. O. Demokritov, Appl. Phys. Lett. \textbf{99}, 162502 (2011).
\bibitem{jungfleisch} M. B. Jungfleisch, A. V. Chumak, V. I. Vasyuchka, A. A. Serga, B. Obry, H. Schultheiss, P. A. Beck, A. D. Karenowska, E. Saitoh, and B. Hillebrands, Appl. Phys. Lett. \textbf{99}, 182512 (2011).
\bibitem{ando-prl} K. Ando, and E. Saitoh, Phys. Rev. Lett. \textbf{109}, 026602 (2012).



\bibitem{chaos} P. E. Wigen, \textit{Nonlinear Phenomena and Chaos in Magnetic Materials} (World Scientific, Singapore 1994).
\bibitem{anderson-suhl} P. W. Anderson and H. Suhl, Phys. Rev. \textbf{100}, 1788 (1955).
\bibitem{fetisov} Y. K. Fetisov, IEEE Trans. Magn. \textbf{35}, 4511 (1999).
\bibitem{prabhakar} A. Prabhakar and D. D. Stancil, J. Appl. Phys. \textbf{85}, 4859 (1999).

\bibitem{zhang} Y. T. Zhang, C. E. Patton, IEEE Trans. Magn. \textbf{22}, 993 (1986).
\bibitem{zhang-jap} Y. T. Zhang, C. E. Patton, and G. Srinivasan, J. Appl. Phys. \textbf{63}, 5433 (1988).
\bibitem{chen} M. Chen, C. E. Patton, G. Srinivasan, and Y. T. Zhang, IEEE Trans. Magn. \textbf{25}, 3485 (1989).


\bibitem{iguchil} R. Iguchil, K. Ando, T. An, E. Saitoh, and T. Sato, IEEE Trans. Magn. \textbf{48}, 3051 (2012).

\bibitem{lustikova} J. Lustikova, Y. Shiomi, Z. Qiu, T. Kikkawa, R. Iguchi, K. Uchida, and E. Saitoh, J. Appl. Phys. \textbf{116}, 153902 (2014).


\bibitem{anderson} E. E. Anderson, Phys. Rev. \textbf{134}, A1581 (1964).



\bibitem{sandweg} C. W. Sandweg, Y. Kajiwara. A. V. Chumak, A. A. Serga, V. I. Vasyuchka, M. B. Jungfleisch, E. Saitoh, and B. Hillebrands, Phys. Rev. Lett. \textbf{106}, 216601 (2011).

\bibitem{castel} V. Castel, N. Vlietstra, B. J. van Wees, and J. Ben Youssef, Phys. Rev. B \textbf{86}, 134419 (2012).



\end{thebibliography}
\end{document}